\documentstyle[12pt]{article}
\title{\nopagebreak
\begin{flushright}
\tenrm UCTP101.02
\end{flushright}\vskip .7in
\large \bf  Non-Vanishing Cosmological Constant $\Lambda$, \\ 
            Phase Transitions, \\ And \\
            $\Lambda$-Dependence Of High Energy Processes}
\author{Freydoon Mansouri\thanks{e-mail address: mansouri@uc.edu}
\\ 
\it \small \it
Physics Department, University of
Cincinnati, Cincinnati, OH 45221}

\date{}

\begin{document}

\maketitle

\begin{abstract}
It is pointed out that a collider experiment involves a local
contribution to the energy-momentum tensor, a circumstance which
is not a common feature of the current state of the Universe at
large
characterized by the cosmological constant $\Lambda_0$. This
contribution may be viewed as a change in the sturcture of
space-time from its large scale form governed by
$\Lambda_0$ to one governed by a $\Lambda$ peculiar to the energy
scale of the experiment. Possible consequences of this effect are
explored by exploiting the asymptotic symmetry of space-time for
non-vanishing $\Lambda$ and its relation to vacuum energy.
\end{abstract}

\pagebreak

\section{introduction}
The notion of phase transition plays
an important role in our understanding of particle physics at the
fundamental level. It also plays an important role in the
inflationary cosmology. One may therefore view this notion 
 as an important factor in shaping our ideas of the
Universe today. Its most concrete realization has been through
spontaneously broken symmetry. For whatever dynamical reasons,
when a phase transition takes place,
there will be a contribution
to the vacuum energy density $\rho_{vac}$ of space-time. On the
basis of Einstein's field equations, such a non-vanishing
vacuum
energy is equivalent to having a cosmological constant,
$\Lambda$, given
by~\cite{weinberg,carroll}
\begin{equation}
\Lambda = 8 \pi G_N \rho_{vac},
\end{equation}
where $G_{N}$ is Newton's constant.
This relation raises a number of interesting issues. Among these,
the one that has received a good deal of attention in recent
years is that, the contribution to $\Lambda$ from
each phase transition, such as that at the electroweak scale, is
very
large~\cite{weinberg,carroll} according to the current theories.
It follows that unless a delicate
cancellation mechanism such as supersymmetry has been in
operation over the entire range of scales from the time of big
bang to present, it would be difficult to account for the
smallness of $\Lambda_{observed}$ according to standard
interpretations. On the other hand, if we view the vacuum energy
arising from
phase transitions as the primary source of
$\Lambda$, a constant $\Lambda$ is not compatible with such phase
transitions.

Another issue which arises from the equality given by Eq. (1)
is that for finite $\Lambda$, the asymptotic symmetry group of 
space-time is not Poincar\'e but de Sitter (dS) or anti-de Sitter
(AdS)
group. On the other hand, the well known Einstein mass-energy
relation $E = mc^2$ is a consequence of Poincar\'e symmetry and
is
valid only in Minkowski space-time. One of the aims of the
present work is to point out how the rest energy expression is
modified in dS$_4$ and AdS$_4$ spaces and how they depend on the
cosmological constant and the vacuum energy. As mentioned above,
if we
take the connection between $\Lambda$ and phase transitions
seriously, there must have been some periods in
the history of the Universe, at which the value of $\Lambda$
changed significantly because of phase transitions. In
particular, at some such periods, the value of $\Lambda$ must
have been very large. It would be of interest to see how the
kinematics of high energy reactions were modified during such
periods and whether this led to any testable consequences. It
will be seen below that if again we take the phase transition, or
its realization as spontaneously broken symmetry, as the
primary
source of a non-vanishing $\Lambda$, it may be possible to test
some of the consequences at the next generation of high energy
colliders. To accomplish this, it will be necessary to reexamine
the relation between the
processes in the current high energy experiments and the period
(energy scale) in the history of the Universe at which such
processes occurred routinely. The
consequences will be significant~\cite{fm} if a high energy
experiment modifies
the local
structure of space-time for a short period of time, so that the
immediate neighborhood of a collision departs from the Minkowski
space and
becomes a dS or AdS space-time. In both dS and AdS spaces the
ground state energies, i.e., the analogs of the energy in
Einstein's mass-energy relation, depend not only on the mass but
also on spin and on $\Lambda$, with the correct value of
$\Lambda$ to be determined such that it is appropriate to the
relevant energy scale.

The mathematical frameworks for obtaining the analogs of
Einstein's mass-energy relation in AdS$_4$ and dS$_4$ spaces
depend on a mixture of the representation theory of the
corresponding groups and the field theoretic constructions in
these spaces. Most of this already exist in the literature in one
form or another. The construction of the representations for
these groups relies on the structure of appropriate little
groups. For AdS$_4$ group, the relevant
little group is~\cite{gursey,fronsdal,evans} the compact
group $SO(2) \times SO(3)$, and it allows for a simple
identification of an appropriate energy operator.
For the dS$_4$ group, the simplest choice for a subgroup is the
maximal compact subgroup~\cite{thomas} $SO(4)$. However, when the
representations are induced in this way, the direct
identification of one of the generators with an energy operator
does
not lead to a physically suitable energy spectrum. So, we will
use an alternative realization of the algebra due to
Bohm~\cite{bohm}. In both
spaces, to relate the
eigenvalues of the quadratic Casimir operators to the physical
masses, it is necessary to introduce the concept of mass
as the pole of the propagator in an appropriate quantum field
theory. To lowest order this pole would be that of the Green's
function for a dS$_4$ ( or AdS$_4$ ) invariant differential
equation. As is familiar from field theories in Minkowski space
time, it is desirable that such an equation become conformally
invariant in the
limit of vanishing mass~\cite{tagirov,isham}. Combining these
concepts, and assuming the existence of an invariant vacuum
state~\cite{tagirov,mottola,allen}, it is then possible to obtain
an expression for the ground state energy. In carrying out this
project, one must also carefully keep track of the dependence on
the cosmological constant, $\Lambda$.
In the discussions of the
representation theories of AdS and dS groups mentioned above, the
explicit values
of the radii of curvatures of the corresponding spaces play no
essential roles, so that they are often set equal to unity. In
view of Eq.
(1) and the point of view emphasized in this work, to extract
physical consequences it is essential that we explicitly display
the dependence on this dimensional parameter or, equivalently,
the
cosmological constant. 

The change in the asymptotic structure of space-time from
Minkowski to dS or AdS spaces implies a change in the asymptotic
symmetry group of space-time. In particular, it implies the
breakdown of translational invariance. In the last decade,
possible violations of Lorentz and CPT symmetries have been
studied in great detail~\cite{alan}. In the same vein, the
results presented in this work may also be viewed as probing the
possible violation of translational invariance.

\section{The dS and AdS Spaces and Algebras}
We begin with some of the well known properties of these spaces
and algebras. The dS and AdS spaces and algebras have many
features in common. So, we
give the details for one of them
and then indicate the changes, if any, in the
corresponding expressions for the other space and algebra.   
As much as possible, we will follow the notation and conventions
of reference~\cite{fm1}.

The anti-de Sitter space in 3+1 dimensions can be viewed  as a
subspace of a
flat 4-dimensional space with the line element
\begin{equation}
ds^2 = dX_A dX^A = dX_0^2 - dX_1^2 -dX_2^2 - dX_3^2 + dX_4^2
\end{equation}
It is determined by the constraint
\begin{equation}
(X_0)^2 - (X_1)^2 - (X_2)^2 - (X_3)^2 + (X_4)^2 = l^{2}
\end{equation}
where $l$ is a real constant and is related to the cosmological
constant according to $\Lambda = - l^{-2}$. The set of
transformations which
leave the line element invariant
form the AdS$_4$ group $SO(3,2)$. We will actually deal with the
universal cover of this group.

Similarly, the dS$_4$ space can be viewed as a subspace of a flat
4-dimensional space with line element
\begin{equation}
ds^2 = dX_A dX^A = dX_0^2 - dX_1^2 -dX_2^2 - dX_3^2 - dX_4^2
\end{equation}
It is determined by the constraint
\begin{equation}
(X_0)^2 - (X_1)^2 - (X_2)^2 - (X_3)^2 - (X_4)^2 = -l^{2},
\end{equation}
where now $\Lambda = l^{-2}$.

The AdS$_4$ algebra consists of the elements $M_{AB}$ satisfying
the
commutation relations
\begin{equation}
[M_{AB}, M_{CD}] = i\left(\eta_{AD} M_{BC} + \eta_{BC}M_{AD} -
\eta_{AC} M_{BD} - \eta_{BD} M_{AC}\right).
\end{equation}
With the index $A=(\mu ,4)$ and $\mu = 0,1,2,3$, we can write the
algebra
in a more familiar four dimensional notation by setting
\begin{equation}
M_{\mu 4} = l \Pi_{\mu}.
\end{equation}
With the notation
\begin{equation} 
\epsilon^{0123} = 1 ; \hspace{1.0cm} \eta^{ab} = (1, -1, -1,
-1),
\end{equation}
we get
$$[M_{\mu \nu}, M_{\rho \sigma}] = i\left(\eta_{\mu \sigma} 
M_{\nu \rho} + \eta_{\nu \rho} M_{\mu \sigma} -
\eta_{\mu \rho} M_{\nu \sigma} - \eta_{\nu \sigma} 
M_{\mu \rho}\right)$$
\begin{equation}
[ M_{\mu \nu}, \Pi_{\rho} ] = i\left(\eta_{\nu \rho} \Pi_{\mu} -
\eta_{\mu \rho} \Pi_{\nu}\right)
\end{equation}
$$[\Pi_{\mu}, \Pi_{\nu}] = - i l^{-2} M_{\mu \nu} = i \Lambda
M_{\mu \nu}.$$
As expressed in terms of the cosmological parameter $\Lambda$,
this algebra also holds for dS$_4$.
It can be seen that, under suitable conditions, in the limit of
vanishing $\Lambda$, the 
AdS$_4$ and dS$_4$ algebras contract to the Poincar\'e algebra.

The AdS$_4$ and dS$_4$ groups have each two Casimir invariants.
The quadratic ones
are given, with appropriate signs for $\Lambda$, by
\begin{equation}
I_1 = - \frac{\Lambda}{2} M_{AB} M^{AB} = \Pi_{\mu} \Pi^{\mu} -
\frac{\Lambda}{2} M_{\mu \nu} M^{\mu \nu}
\end{equation}
The important
point here is that, in contrast to Poincar\'e algebra, for
AdS$_4$ as well as dS$_4$ algebras the
quantity $\Pi_{\mu}
\Pi^{\mu}$ is not an invariant by itself. As a result, the
analogs of
Einstein mass-energy relation $E = mc^2$ will also depend on
spin and the cosmological constant, as we shall see below.

The second Casimir invariant $I_2$ for either AdS$_4$ or dS$_4$
algebras can be
written as
\begin{equation}
I_2 = \frac{\Lambda}{16} V_A V^A,
\end{equation}
where
\begin{equation}
V^A = \epsilon^{ABCDE} M_{BC} M_{DE}.
\end{equation}
Defining the analog of the Pauli-Lubanski operator as
\begin{equation}
W^{\mu} = V^{\mu} = \epsilon^{\mu \nu \rho \sigma } \Pi_{\nu} 
M_{\rho \sigma},
\end{equation}
we can write $I_2$ in four dimensional notation for both
algebras:
\begin{equation}
I_2 = W_{\mu} W^{\mu} - \frac{\Lambda}{16} (V^4)^2
\end{equation}

\section{Massive Unitary Representations}
It will be recalled that for the Poincar\'e group, the
construction of unitary representations begins by identifying an
appropriate little group. For massive momenta, e.g., the little
group is $SO(3)$. Then the corresponding induced
representation would be specified by the ground state eigenvalues
of the Casimir operators,
i.e., by the (rest) energy $E_0$ and spin $s$. The
quantity $E_0^2$ is thus the ground state eigenvalue of the
quadratic Casimir
operator $P_{\mu} P^{\mu}$ of the Poincar\'e algebra. The notion
of mass is then introduced as the pole of the propagator in the
corresponding field theory. For the Poincar\'e group, this
equation is the
relativistically invariant free massive Klein-Gordon equation.
Then, noting that 
the d'Alembertian operator is just the representation of the
Casimir operator $P_{\mu} P^{\mu}$ in terms of differential
operators, one obtains:
\begin{equation}
[ P_{\mu} P^{\mu} - m^2 c^4 ] \phi = 0
\end{equation}
This
provides a group theoretic basis for obtaining the expression
$E_0 = mc^2$. One can follow the same recipe for relating the
notions of
mass and energy
in dS$_4$ and AdS$_4$ spaces. To this end, we must first obtain
the ground state eigenvalues of the corresponding quadratic
Casimir operators. We begin with AdS$_4$.

The massive discrete unitary representations for AdS$_4$
algebra
has been known for sometime~\cite{fronsdal,evans}. With $SO(2)
\times SO(3)$ as maximal compact subgroup, let
\begin{equation}
H = l \Pi_0; \quad J_{\pm} = M_{23} \pm i M_{31}; \quad J_3 =
M_{12}.
\end{equation}
The $J$'s form the algebra of the $SO(3)$. They commute with $H$
which generates the $SO(2)$ part.
The quantity $\Pi^{\mu}$ has the same dimension as $P^{\mu}$ in
the Poincar\'e algebra, so that it is natural to identify $\Pi^0$
as the energy operator. The generator $H$ is then the
dimensionless version of the energy operator. So, we can label
our states by the eigenvalues of one or the other of these
operators.

We can choose the remaining six generators of AdS$_4$ algebra
such that they can act as raising and lowering operators for
eigenvalues of $\Pi^0$. To this end, let
\begin{equation}
B_i^{\pm} = M_{0i} \pm i l \Pi_i  ; \quad i=1,2,3.
\end{equation}
All three of the plus (minus) operators raise (lower) the
eigenvalues of $H$ :
\begin{equation}
[ H, B_i^{\pm} ] = \pm B_i^{\pm}.
\end{equation}
In this $\{ H, J, B \}$ basis, the quadratic Casimir operator
will take the form
\begin{equation}
I_1 = \Pi_0 \, ( \, \Pi_0 - \frac{3}{l} \, ) + \frac{1}{l^2}
\vec{J}^2 
+ \frac{1}{l^2} \, B_i^+ \, B_i^- ,
\end{equation}
where in the last term a sum over the index ``i '' is to be
taken.

Using the above preparations, we can now
label~\cite{fronsdal,evans}
the discrete representations of AdS$_4$ algebra by
the
labels induced by subgroup $SO(2)
\times SO(3)$.
Depending on whether we use eigenvalues of $\Pi_0$ or $H$, they
are:
\begin{equation}
| E, j, j_3, E_{\Lambda} > \, ; \quad |h, j, j_3, E_{\Lambda} >.
\end{equation}
The quantity $E_{\Lambda}$ is the energy scale associated with
the value of $\Lambda$ or, equivalently, with the radius of
curvature
$l$.
For these unitary representations, we denote the lowest (ground
state) eigenvalues by $E_0$ and
$s$, respectively. This means, in particular, that
\begin{equation} 
B_i^- | E_0, s, s_3, E_{\Lambda} > = 0.
\end{equation}
Applying the quadratic Casimir operator on a ground state, we get
\begin{equation}
I_{AdS} | E_0, s, s_3, E_{\Lambda} > = [ E_0 (E_0 -3 E_{\Lambda}
) +
E_{\Lambda}^2 s ( s+1) ] | E_0, s, s_3, E_{\Lambda} >,
\end{equation}
Comparing this
expression with the corresponding eigenvalue for the Poincar\'e
algebra, it is already clear that the ground state energy $E_0$
will depend on $s$ and on
$E_{\Lambda}$.

Next, consider the dS$_4$ algebra. In this case, the maximal
compact subgroup is $SO(4)$. Its labels can be used to induce and
classify the unitary representations of the dS$_4$
algebra~\cite{thomas}. To construct an appropriate 
energy-momentum operator, however, we will make use of a
representation
of dS$_4$ algebra pioneered by Bohm~\cite{bohm}, in which the
unitary principal series of dS$_4$ are realized on an identical
pair of positive
energy representations of the Poincar\'e group. Both the
Poincar\'e and dS$_4$ groups contain the Lorentz group as a
subgroup. The corresponding generators, $M_{\mu \nu}$ satisfy the
commutation relations given in Eq. (9). Let $P_{\mu}$ represent
the (commuting) translation generators of the Poincar\'e algebra,
so that its quadratic Casimir operator is given by $P^2 = 
P_{\mu} P^{\mu}$. Then, using a notation similar to that for
AdS$_4$ given above, let
\begin{equation}
\Pi_{\mu} = P_{\mu} + [ \frac{\Lambda}{2 P^2} ]^{1/2} \{ P^{\nu},
M_{\nu \mu} \},
\end{equation}
where, as in previous section, $\Lambda = l^{-2}$ is the
cosmological constant. It is easy to check that the $\Pi_{\mu}$'s
defined in this way satisfy the commutation relations given in
Eq. (9). Therefore, we have a representation of the dS$_4$
algebra in terms of the operators $P_{\mu}$ and $M_{\mu \nu}$.

In this representation of dS$_4$ algebra, the quadratic Casimir
operator of the algebra given by Eq. (10) can be expressed as
\begin{equation}
I_1 = P_{\mu} P^{\mu} + \frac{9}{4} E_{\Lambda}^2 +
E_{\Lambda}^2 \, \omega_{\mu} \omega^{\mu}.
\end{equation}
In this equation, the quantity 
\begin{equation}
\omega_{\mu} = \frac{1}{2} \epsilon_{\mu \nu \rho \sigma} P^{\nu}
M^{\rho \sigma}
\end{equation}
is the Pauli-Lubanski operator of the Poincar\'e algebra. Also,
as in the case of AdS$_4$ algebra, we have replaced the
cosmological constant with the corresponding energy scale
$E_{\Lambda}$.

We can now compute the eigenvalues of $I_1$ acting on positive
energy representations of the Poincar\'e group. We label the
Poincar\'e state with its ground state (rest) energy $E_0$ and
its spins $s$. Thus we have
\begin{equation}
I_{dS} | E_0, s, E_{\Lambda} > = [ E_0^2 + E_{\Lambda}^2 (
\frac{9}{4} -
 s^2 -s )] | E_0, s, E_{\Lambda} >.
\end{equation}
The range of values of $s$ is the same as one of labels for the
unitary principal series representations of the dS$_4$ group, so
that we
can take $s$ as a suitable definition of spin in dS$_4$ space.
Similarly, the operator $P^0$ has the correct dimension and the
range of eigenvalues to be identified as the energy operator. We
note, however, that, in contrast to Poincar\'e algebra, since
$P_{\mu} P^{\mu}$ is not an invariant of dS$_4$ algebra, $E_0
\not= m c^2$ in dS$_4$ space. The connection to mass will be
discussed in the next section. 

\section{$\Lambda$-Dependence of Mass-Spin-Energy Relations}
We now turn to the issue of how the notion of mass arises in
dS$_4$ or AdS$_4$ spaces and
how it gets related to the eigenvalues of their respective
quadratic Casimir operators. In a standard field theory, the mass
of a particle is identified as the pole of the propagator for the
corresponding field. To lowest order, the propagator is the
classical Green's function for an invariant differential
equation. For a scalar field in Minkowski space, e.g., it is the
Green's function for the Klein-Gordon equation
\begin{equation}
\left( \partial^{\mu} \partial_{\mu} + m^2 \right) \phi (x) =
0,
\end{equation}
This equation immediately generalizes to curved space-time by
replacing $\partial_{\mu}$ with an appropriate covariant
derivative $\nabla_{\mu}$:
\begin{equation}
\left( \nabla_{\mu} \nabla^{\mu} + m^2 \right) \phi (x) = 0.
\end{equation}
It has been observed, however, that in contrast to Klein-Gordon
equation the massless limit of this equation is not conformally
invariant~\cite{tagirov,isham}.
If we require, in analogy with the corresponding situation in
Minkowski space, that the massless limit of the
relevant differential equation be conformally invariant, we must
modify this equation by a term proportional to the scalar
curvature~\cite{tagirov,isham}:
\begin{equation}
\left( \nabla_{\mu} \nabla^{\mu} + \frac{R}{3} + m^2 \right) 
\phi (x)
= 0.
\end{equation}
where for dS$_4$ and AdS$_4$ spaces,
\begin{equation}
R = \pm \frac{3}{2 l^2}
\end{equation}
More generally, for a particle of any spin, we have
\begin{equation}
\left( \nabla_{\mu} \nabla^{\mu} - E_{\Lambda}^2 \, \beta_s + 
m^2 c^4 \right) \Psi_s = 0,
\end{equation}
where $\beta_s$ is a rational number that can be different for
different spin. It is to be noted that the question of whether or
not the massless limit of a wave equation in curved space should
have the same symmetry, i.e., conformal invariance, as that of
the corresponding equation in Minkowski space, must ultimately be
settled by its physical predictions. As far as the result given
below are concerned, conformal invariance or lack thereof affects
the value of the coeffeciant $\beta_s$.

With a suitable differential equation at hand, it is now straight
forward to connect it to the quadratic Casimir operator, $I_1$,
of the dS$_4$ or AdS$_4$ algebra. This can be worked out in a
manner which
is familiar from the relation indicated in the previous section
between the
d'Alembertian operator and the quadratic Casimir operator of the
Poincar\'e (Lie) algebra in Minkowski space. This means that to
have an invariant differential equation, we must have 
\begin{equation}
I_1 = - \nabla_{\mu} \nabla^{\mu} + b(s).
\end{equation}
where the quantity $b(s)$ can depend on spin. Using the field
equation and the eigenvalues for $I_1$ of dS$_4$ given by Eq.
(26), we get
\begin{equation}
m^2 c^4 = E_0^2 + 
E_{\Lambda}^2 ( \frac{9}{4} - s^2 - s + \beta_s ) - b(s).
\end{equation}
Similarly, using Eq. (22) for AdS$_4$, we get, with in general a
different value for $\beta_s$,
\begin{equation}
m^2 c^4 = E_0 ( E_0 - 3 E_{\Lambda} ) +
E_{\Lambda}^2 ( s^2 + s + \beta_s ) - b(s).
\end{equation}
The precise form of $b(s)$ will depend on the choice of unitary
representations and other constraints. For example, for the
discrete unitary series of AdS$_4$, one can use the
non-negativity of the
norms of the states~\cite{fronsdal} to get
\begin{equation}
m^2 c^4 = E_0 ( E_0 - 3 E_{\Lambda} ) + 
E_{\Lambda}^2 \, ( 2 + s - s^2 ).
\end{equation}
What is important for our purposes
is that both for dS$_4$ and AdS$_4$ the expression for the ground
state energy is
of the form
\begin{equation}
E_0 = E_0 (m, s, E_{\Lambda}).
\end{equation}

\section{Potential Experimental Consequences and Speculations}
One of the straight forward but important consequences of a 
non-vanishing cosmological constant is that it changes the
asymptotic
symmetry
group of space-time from Poincar\'e group to de Sitter or
anti-de Sitter group. One immediate consequence of this is that
the familiar Einstein rest energy expression $E_0 = m c^2$
is modified to an expression given by Eq. (36), involving mass,
spin, and an energy
scale determined by the cosmological constant. Moreover, if our
current theories of particle physics based on spontaneously
broken symmetry and phase transition are correct, then there must
have been some periods in the history of the Universe, in which
the value of $\Lambda$, and hence of $E_{\Lambda}$, changed
significantly. This is clearly not compatible with the idea that
a single constant $\Lambda$, vanishing or non-vanishing, has
controlled the expansion of the Universe. It thus appears that a
suitable theoretical framework would have to regard $\Lambda$ as
a dynamical quantity, which is both space and time dependent.
The explicit construction of such a framework is beyond the scope
of this work. Here, we confine the discussion to some general
remarks on the $\Lambda$-dependence of high energy processes.

Let us consider one of the potential experimental consequences of
the
expression given by Eq. (36). To this end, it will be recalled
that the quantity
$E_{\Lambda}$ is the energy associated with the radius of
curvature, $l_{\Lambda}$, of the AdS$_4$ or dS$_4$ space. This
energy scale is related to but is distinct from the scale at
which the vacuum energy is evaluated. The difference between the
two has to do with the appearance of Newton's constant $G$
in Eq. (1). As a first try, let
us assume, somewhat naively, that it is the current
value $\Lambda_0$ of the cosmological parameter that should set
the energy
scale $E_{\Lambda}$. Then, given the current
bounds~\cite{weinberg,carroll}, 
\begin{equation}
l_{\Lambda_{0}} = | \Lambda_{0} |^{- \frac{1}{2}} \sim
10^{28} cm.
\end{equation}
It follows that
\begin{equation}
E_{\Lambda_0} \sim 10^{- 33} eV
\end{equation}
So, if the relevant energy scale $E_{\Lambda}$ in
Eq. (36) were set by the current value of the cosmological
parameter, the deviation from $E = m c^2$ would be very small,
and
there would be no hope for the experimental detection of such
deviations in the existing or the planned future colliders. 

On the other hand, if our current theories of particle physics
based on spontaneously broken symmetry and phase transition are
correct, then there must have been some periods in the history of
the Universe, in which the value of $\Lambda$, and hence of
$E_{\Lambda}$, were large. For example, in the
electroweak epoch characterized by $\Lambda_{EW}$, the kinematics
of a typical electroweak process must have depended significantly
on $E_{\Lambda_{EW}}$. It is then of interest to see whether this
kind of $\Lambda$-dependence can be detected in high energy
collicer experiments. In this respect, we note that a typical
high energy collider experiment at, say, 20 TeV does not
necessarily involve a
phase transition leading to a significantly larger value of
$\Lambda$. But it is conceivable that in future collider
machines a mechanism could be devised such that, in the
immediate neighborhood of the collision, one can justify the
connection between the scale of the experiment and the value of
$\Lambda$ corresponding to a phase transition at that scale.
Assuming such
a connection, let us compute the expected order of magnitude
deviations from $E = m c^2$ law.

For energies of the order of 200 GeV corresponding to
the electroweak phase transition, the radius of curvature
$l_{\Lambda_{EW}} \sim \frac{1}{4}$ cm. So, 
$E_{\Lambda_{EW}}\sim 10^{-4}$ eV.
For high energy experiments of order 20 TeV, one finds
$E_{\Lambda_{TeV}} \sim 1$ eV. Finally, for energies of order
1000 TeV, the value of $E_{\Lambda}\sim 2500$ eV. For
particles of small mass such as neutrinos, these deviations from
the standard rest energy will lead to
significant changes in kinematics at very high energies.
It may be possible to test
this proposal in not
too distant a future.

\vspace{1.5cm}

\noindent{\bf Acknowledgments:}
This work was supported, in part by the Department of Energy
under the contract number DOE-FGO2-84ER40153. I would like to
thank Philip Argyres for reading the manuscript and making
valuable suggestions for improvement. I am also indebted to
Professor Arno Bohm for discussion and correspondence concerning
his work.

\end{document}